# Unseen Attack Detection in Software-Defined Networking Using a BERT-Based Large Language Model

Mohammed N. Swileh (mohammedswileh2023@email.szu.edu.cn)[a,1], Shengli Zhang (zsl@szu.edu.cn)[a,*]

[a] *College of Electronics and Information Engineering, Shenzhen University, Shenzhen, 518060, China*

[1] *First author*

[*] *Corresponding author*

## Abstract

Software-defined networking (SDN) represents a transformative shift in network architecture by decoupling the control plane from the data plane, enabling centralized and flexible management of network resources. However, this architectural shift introduces significant security challenges, as SDN's centralized control becomes an attractive target for various types of attacks. While current research has yielded valuable insights into attack detection in SDN, critical gaps remain. Addressing challenges in feature selection, broadening the scope beyond DDoS attacks, strengthening attack decisions based on multi-flow analysis, and building models capable of detecting unseen attacks that they have not been explicitly trained on are essential steps toward advancing security in SDN. In this paper, we introduce a novel approach that leverages Natural Language Processing (NLP) and the pre-trained BERT-base model to enhance attack detection in SDN. Our approach transforms network flow data into a format interpretable by language models, allowing BERT to capture intricate patterns and relationships within network traffic. By Using Random Forest for feature selection, we optimize model performance and reduce computational overhead, ensuring accurate detection. Attack decisions are made based on several flows, providing stronger and more reliable detection of malicious traffic. Furthermore, our approach is specifically designed to detect previously unseen attacks, offering a solution for identifying threats that the model was not explicitly trained on. To rigorously evaluate our approach, we conducted experiments in two scenarios: one focused on detecting known attacks, achieving 99.96% accuracy, and another on detecting unseen attacks, where our model achieved 99.96% accuracy, demonstrating the robustness of our approach in detecting evolving threats to improve the security of SDN networks.

## 1. Introduction

The rapid evolution of digital ecosystems demands networks that can efficiently adapt to changing conditions, yet conventional networks often struggle with intrinsic limitations that hinder their ability to meet modern requirements. These networks are constrained by their rigid and static architectures, where tightly coupled network devices limit scalability and agility, as stated by (Kreutz et al. 2014). Furthermore, decentralized management in traditional networks introduces inconsistencies, inefficiencies, and challenges in enforcing policies, as highlighted by (Swami et al. 2023). The absence of programmability and automation further exacerbates these issues, making it difficult to quickly deploy and manage network services, optimize traffic routing, or ensure Quality of Service (QoS). This lack of centralized control impairs visibility over network traffic, obstructing efforts to efficiently route data, enforce security policies, and optimize network performance.

In contrast, Software-defined networking (SDN) has emerged as a paradigm-shifting technology in the evolving landscape of modern networking, fundamentally altering traditional approaches to network architecture, design, and management (Tang et al., 2023). The core innovation of SDN lies in its ability to separate the control plane from the data plane, consolidating network intelligence and management into software-based controllers (Riggio et al., 2015). This shift enables organizations to achieve unparalleled flexibility, scalability, and agility in overseeing network infrastructure. By abstracting network control into software, SDN empowers administrators to dynamically configure and manage network resources, optimize traffic routing, and implement policies with granular control. Its programmable nature not only simplifies network operations but also accelerates the deployment of new services and applications, fostering continuous adaptability and innovation. Moreover, the centralized control and programmability of SDN deliver numerous benefits, including improved security, enhanced resource utilization, increased network efficiency, and seamless scalability (Xia et al., 2014). As more organizations recognize SDN's transformative potential, its adoption continues to grow across diverse sectors. Notable examples include Google, which has employed SDN in its global network through its Espresso solution since 2012, and AT&T, which has utilized the ECOMP platform to enhance its network management and service delivery since 2014. Similarly, Facebook has used SDN to optimize its global network infrastructure with its Fabric technology since 2011, and NTT Communications has leveraged SDN for its Enterprise Cloud to improve resource allocation and cost efficiency since 2013 (Hussain et al. 2023).

SDN consists of a well-organized architecture with three essential layers: the application layer, the control layer, and the infrastructure layer (Singh and Behal, 2020). The application layer operates as a versatile platform for various network services and applications, encompassing virtualized functions, orchestration tools, and service-chaining techniques. The control layer is positioned between the application and infrastructure layers. It contains the centralized SDN controller, which manages network policies, configurations, and traffic flows using standardized protocols like OpenFlow. This controller facilitates communication with network devices in the infrastructure layer, comprising physical and virtual components such as switches, routers, and NFV components. These devices carry out instructions from the controller, allowing for the transmission of dynamic data, implementation of Quality of Service (QoS), and adaptive routing. SDN distinguishes itself from previous networking approaches by offering a structured design that enables centralized control, programmability, and efficient resource usage. This architecture also promotes flexible, responsive, and scalable network administration.

Despite the numerous benefits of Software-Defined Networking (SDN), its centralized architecture inherently introduces significant vulnerabilities, making it susceptible to a range of security threats across all layers (Benzekki et al., 2016). These threats can be categorized into four main vectors: SDN Controller Attacks, Planes Communication Attacks, Application Plane Attacks, and Data Plane Attacks, as illustrated in Figure 1. The very design of SDN, which separates the data plane from the control plane, gives rise to unique attack vectors specific to SDN architecture such as SDN Controller Attacks, and Planes Communication Attacks (Elsayed et al., 2020). While some threats are also present in traditional networks such as Application Plane Attacks, and Data Plane Attacks, their impact in an SDN environment is often amplified; for instance, unauthorized access in a conventional network may compromise only a single device, whereas in an SDN, it can jeopardize the entire network due to its centralized control. This reality underscores the critical need for robust protection strategies to safeguard SDN networks from diverse and evolving threats, ensuring their resilience and operational integrity.

The centralized control and programmability of SDN not only enhances operational efficiency but also renders it a prime target for a diverse range of malicious actors seeking to exploit network vulnerabilities (Sahoo et al., 2018). Among the most pressing threats are Distributed Denial of Service (DDoS) attacks, which can disrupt fundamental network operations by overwhelming resources with malicious traffic. However, the threat landscape encompasses more than just DDoS; it includes Probe, DoS, Botnet, U2R, Web attacks, and BFA, each presenting unique challenges that could severely compromise SDN environments. DDoS attacks can saturate both the data plane and the SDN controller, while Probe attacks focus on gathering critical information that may pave the way for more severe breaches. DoS attacks specifically target individual services, jeopardizing availability, and Botnet attacks leverage networks of compromised devices to amplify their impact. U2R attacks seek unauthorized access to gain control over network resources, threatening data integrity and confidentiality, while Web attacks exploit vulnerabilities in applications running within the SDN. BFA attempts to gain unauthorized access through repeated login attempts to the controller, posing significant risks to SDN security and potentially allowing attackers to manipulate SDN resources. The centralized architecture, although beneficial for efficient network management, becomes a double-edged sword, transforming it into a potential single point of failure. This reality highlights the urgent need for comprehensive defense mechanisms, proactive threat detection, and swift incident response strategies to protect SDN environments from this evolving array of threats.

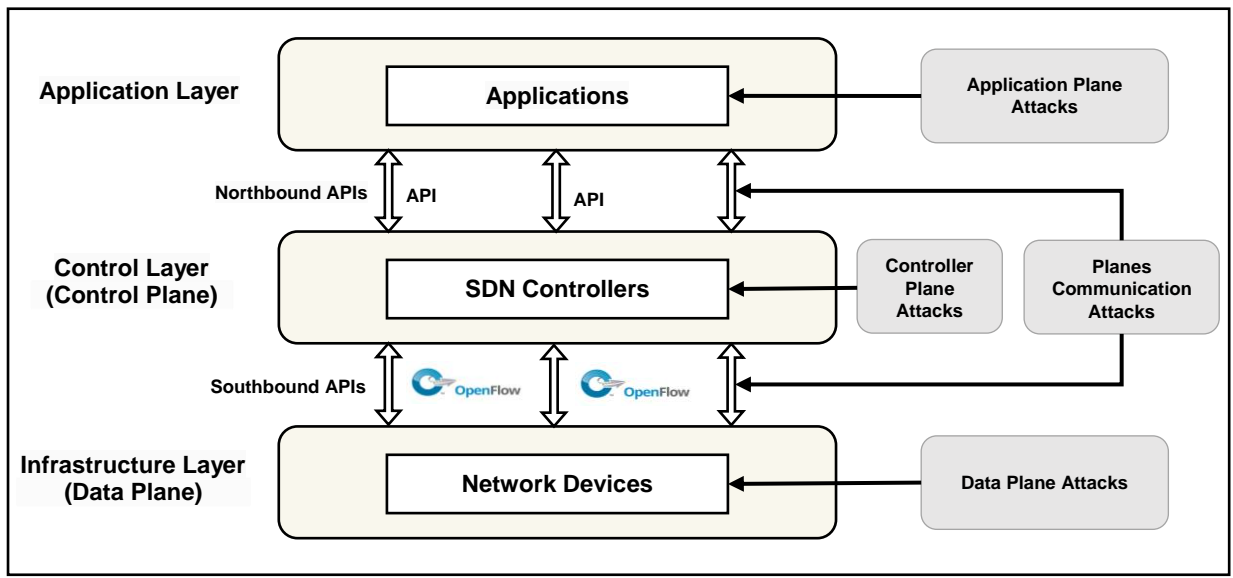

*Figure 1: The Main Attacks Vectors in SDN architecture*

While SDN has witnessed the development of various security techniques, including machine learning (ML) and deep learning (DL) methods for attack detection as reviewed by Singh and Behal (2020), Wabi et al. (2023), Bahashwan et al. (2023), and Ali et al. (2023) these models come with significant limitations. A key challenge lies in their limited ability to capture complex relationships in network traffic, as they often analyze individual flows in isolation, potentially missing important contextual patterns that are critical for detecting sophisticated attack strategies. Furthermore, ML and DL models struggle with adapting to evolving attack vectors, particularly when faced with unseen or zero-day attacks that were not part of the training dataset. This vulnerability diminishes their effectiveness in dynamic, real-world environments. Moreover, the performance of these models is highly sensitive to the quality of the input data, often leading to overfitting when the data is noisy or imbalanced. Scalability also becomes an issue, especially in large SDN environments where the volume of network traffic data is substantial. While training time can be a consideration for all models, the interpretability and adaptability of traditional ML and DL approaches are particularly limited, making it difficult to deploy them confidently in real-world security applications. These limitations underscore the need for more advanced techniques capable of understanding broader contexts in network traffic and detecting both known and unknown attack patterns.

In contrast to traditional ML and DL models, leveraging the capabilities of Natural language processing (NLP) and large language models (LLMs) from the Transformer family offers significant advantages for detecting attacks in SDN networks. Transformers, particularly BERT, excel in capturing contextual information and complex dependencies within network traffic through their self-attention mechanisms, allowing them to identify intricate patterns and relationships that conventional models may overlook. This capability is critical for accurately detecting sophisticated or coordinated attack strategies across multiple network flows. Furthermore, Transformers are pre-trained on vast amounts of data, which allows them to generalize more effectively and reduces the need for extensive feature engineering specific to each task. Their ability to handle unstructured and high-dimensional data makes them adaptable to diverse network traffic patterns, enabling them to detect even subtle anomalies that could signify new or evolving threats. Importantly, Transformers exhibit a robust capacity for detecting unseen attacks, a limitation in traditional ML and DL approaches. By recognizing attack vectors not included in the training datasets, these models improve the security of SDN environments against novel or zero-day attacks. The transfer learning capabilities of Transformers further enhance their value, as they can be fine-tuned on smaller, domain-specific datasets, reducing the need for massive amounts of labeled data while maintaining and improving detection performance. Despite these advantages, researchers have not yet explored the potential of utilizing Transformers for attack detection in SDN environments. This unexplored potential represents a significant opportunity for advancing the accuracy and effectiveness of attack detection by harnessing the advanced capabilities of these models.

To safeguard SDN networks from evolving and sophisticated threats, it is imperative to develop specialized detection approaches that can effectively address the unique challenges posed by modern attacks. In this paper, we present a novel and comprehensive approach designed to enhance SDN security, Against different types of attacks. Our approach integrates advanced feature selection and data transformation into Natural Language sentences to leverage a pre-trained BERT model, creating a more accurate and adaptive detection approach. Our contributions can be summarized as follows:

1. Leveraging BERT-base-uncased for Network Data in Natural Language Format: Our approach harnesses the pre-trained BERT-base-uncased model by transforming network flow data into natural language sentences, allowing BERT to effectively capture and interpret complex traffic patterns for attack detection in SDN environments. This transformation enhances detection accuracy and reduces false positives by leveraging BERT's deep contextual understanding of network behaviors.

2. Feature Selection for Enhanced Performance: To accommodate BERT's limited sentence length, we employ a Random Forest Classifier to select the top ten most important features from the dataset. This targeted selection optimizes model efficiency, reducing computational overhead while maintaining robust attack detection capabilities by focusing on the features most critical for distinguishing normal and malicious traffic.

3. Multi-Flow-Based Attack Detection: Our approach leverages multiple network flows to improve the Performance of attack detection. By analyzing patterns across multiple flows, rather than relying on single-flow observations, it captures complex behaviors associated with various attack types. This multi-flow perspective enables a more robust and reliable decision-making process for detecting attacks in SDN environments.

4. Unseen Attack Detection with BERT: One of the key strengths of our approach is its ability to detect previously unseen attacks. By leveraging the generalization capabilities of the pre-trained BERT model, our approach can identify novel attack vectors and zero-day threats that were not present in the training data, making it highly adaptable to the continuously evolving landscape of network security.

The remainder of this paper is structured as follows. Section 2 reviews the related work, offering a comprehensive overview of advancements and limitations in existing attack detection methods within SDN environments, and highlights the gaps that our research aims to address. Additionally, it outlines the background and underlying algorithm of the BERT-base-uncased model, emphasizing its relevance and potential for improving attack detection in SDN networks. Section 3 introduces our proposed methodology, which includes Data Processing and Feature Selection, the Transformation of the Dataset into NLP format, the Combination of Flows, Dataset Distribution, and the Fine-tuning of the BERT-base-uncased Model. Section 4 presents our Experimental Performance Analysis, detailing the experimental setup, the selected dataset and Results of Features importance, and the Model performance metrics. It includes an in-depth analysis of the model's performance results and a comparison of the BERT model with the DNN and CNN model. Finally, Section 5 concludes the paper by summarizing key findings and providing directions for future research.

## 2. Related work and Background Algorithm of BERT-base-uncased

This section provides a thorough review of related work, offering insights into the current advancements and challenges within SDN attack detection methods. By examining existing approaches, we highlight the limitations that motivate our proposed solution and underscore the need for more effective detection techniques. Additionally, this section presents an overview of the BERT-base-uncased model, detailing its foundational algorithm and discussing its unique advantages for attack detection in SDN environments.

### 2.1 Related work

The evolution of network security has been profoundly influenced by the advent of Software-Defined Networking (SDN), introducing significant opportunities alongside formidable challenges. SDN's centralized control and programmability have redefined network architecture, enhancing flexibility and efficiency in network management. However, this shift has also rendered SDN environments more vulnerable to sophisticated cyber threats, particularly DDoS attacks. As SDN adoption grows, the necessity for robust security mechanisms to combat these threats becomes increasingly critical. This section provides a detailed review of the current literature on attack detection within SDN environments, exploring various methodologies, techniques, and empirical findings from recent studies aimed at fortifying SDN infrastructures against persistent threats. The objective of this review is to synthesize existing knowledge while identifying gaps, challenges, and opportunities for advancing DDoS defense strategies within the dynamic context of SDN. Table 1 presents a comprehensive summary of the latest techniques employed to defend SDN against attacks (Najar et al. 2024; Hnamte et al. 2024; Gadallah et al. 2024; Chouhan et al. 2023; Zainudin et al. 2022; Cil et al. 2021; Perez-Diaz et al. 2020; Elmasry et al. 2020; Said Elsayed et al. 2020), detailing the models utilized, specific datasets, methodologies applied, and key findings associated with each approach, thereby offering a thorough understanding of the diverse strategies developed to counteract attacks.

*Table 1: Summary of Existing literature*

| Author and Year | Model | Dataset | Methodology | Main Findings |
|---|---|---|---|---|
| Najar et al. (2024) | BRS + CNN | CICDDoS2019 | BRS + CNN-based approach utilizing Balanced Random Sampling and CNNs for attack detection in SDN environments. | Achieved 99.99% accuracy in binary classification, and 98.64% accuracy in multi-classification. |
| Hnamte et al. (2024) | DNN | InSDN, CICIDS2018, Kaggle DDoS | A deep neural network (DNN) architecture with supervised learning techniques for DDoS attack detection in SDN environments | Detection accuracy rates of 99.98% (InSDN), 100% (CICIDS2018), 99.99% (Kaggle DDoS) with low loss rates. |
| Gadallah et al. (2024) | AE-BGRU | Generated Dataset, NSL-KDD | Deep Learning model (AE-BGRU) with Selected features for DDoS detection in SDN control and data planes | AE-BGRU achieved high performance with an accuracy of 99.87%, precision of 98.99%, recall of 99.48%, and F-measure of 99.18% |
| Chouhan et al. (2023) | SVM, RF, KNN, XGBoost, NB | Generated Dataset | Utilized feature extraction and SVM, Random Forest (RF), K-Nearest Neighbors (KNN), XGBoost, and Naïve Bayes (NB) classifiers to detect DDoS attacks in SDN. | The SVM classifier outperformed all other classifiers, achieving an accuracy of 99.398%, precision of 99.413%, recall of 99.397%, an AUC of 0.995, and an F1 score of 99.400% for DDoS attack detection in SDN. |
| Zainudin et al. (2022) | Hybrid (CNN-LSTM) | CICDDoS2019 | Utilized XGBoost for feature selection and a hybrid CNN-LSTM model for DDoS attack classification in SDN-based IIoT networks. | Achieved 99.50% accuracy in DDoS attack classification in SDN-based IIoT networks using XGBoost-based feature selection and a hybrid CNN-LSTM model, demonstrating low-complexity capability and high accuracy for low-latency requirements. |
| Cil et al. (2021) | DNN | CICDDoS2019 | Utilized a deep neural network (DNN) for DDoS attack detection and classifying attack types | Attained 99.99% success in detecting DDoS attacks and achieved 94.57% accuracy in classifying attack types |
| Perez-Diaz et al. (2020) | J48, RT, REP Tree, RF, MLP, SVM | CICDD0S2017 | Proposed a flexible modular architecture for LR-DDoS detection using six ML models (J48, Random Tree, REP Tree, Random Forest, MLP, SVM) in an SDN environment. | The top performer Achieved MLP with an accuracy of 95.00%, precision of 95.46%, recall of 94.51%, and F1-measure of 94.98%. |

| Elmasry et al. (2020) | DNN, LSTM-RNN, DBN | NSL-KDD, CICIDS2017 | Utilized a double Particle Swarm Optimization (PSO)-based algorithm for feature and hyperparameter selection. Employed three deep learning models: DNN, LSTM-RNN, and DBN. | The top performer Achieved DBN. On the CICIDS2017 dataset, it achieved an accuracy of 99.91%, precision of 99.99%, recall of 99.92%, and F1-measure of 99.95%. On the NSL-KDD dataset, it attained an accuracy of 99.79%, precision of 99.83%, recall of 99.81%, and F1-measure of 99.82%. |
|---|---|---|---|---|
| Said Elsayed et al. (2020) | LSTM & OC-SVM | InSDN | LSTM autoencoder combined with One-class Support Vector Machine (OC-SVM) | Achieved an accuracy of 90.5%, precision of 93%, recall of 93%, and F1-measure of 93%, demonstrating the model's effectiveness in detecting anomalies in unbalanced datasets. |

Despite the significant advancements in attack detection within SDN, several gaps in the existing literature highlight areas that warrant further exploration. A recurring issue in many of these studies is the lack of proper feature selection techniques. Several approaches rely on a broad set of features without a clear strategy to isolate those most relevant to attack detection such as studies like those by Najar et al. (2024), Hnamte et al. (2024), Cil et al. (2021), and Perez-Diaz et al. (2020). The use of excessive or irrelevant features not only increases the computational overhead but also introduces the risk of overfitting, which can degrade the model's ability to generalize across different datasets or real-world network environments. By not narrowing down the feature space to those that offer the most significant predictive power, these studies potentially compromise both the efficiency and effectiveness of their models.

Moreover, a predominant focus on DDoS attack detection is evident throughout much of the literature such as studies like those by Najar et al. (2024), Hnamte et al. (2024), Gadallah et al. (2024), Chouhan et al. (2023), Zainudin et al. (2022), Cil et al. (2021), Perez-Diaz et al. (2020), Elmasry et al. (2020) and Said Elsayed et al. (2020), leaving other types of attacks largely unexplored. While DDoS attacks are a well-known and pressing concern in SDN environments, the scope of security threats extends far beyond this singular category. Attack types of threats pose substantial risks, yet they have not been adequately addressed by current methodologies. This narrow focus limits the applicability and robustness of the models when deployed in diverse and dynamic network conditions where multiple forms of attacks may occur simultaneously. Consequently, there is a pressing need for research that adopts a more comprehensive approach to threat detection, accounting for the full spectrum of potential attacks within SDN.

Additionally, all of the existing approaches do not leverage multiple network flows for detecting attacks, which could significantly enhance detection accuracy. By relying solely on a single flow for attack classification, these models miss out on valuable contextual information that could be derived from analyzing previous flows. In real-world scenarios, network interactions are often more complex, and the use of multiple flows of information could provide a richer basis for detecting sophisticated attack patterns. This limitation presents a critical weakness in current detection systems, particularly as network environments grow in scale and complexity.

Another notable gap in the literature is the lack of attention given to the development of models capable of detecting unseen attacks those for which the model has not been explicitly trained. As cyber threats evolve, attackers are constantly developing new techniques that differ from known attack signatures. However, current models tend to focus on detecting only known attack types, leaving SDN networks vulnerable to novel threats. This gap underscores the need for the development of more adaptive models that can generalize to previously unseen attacks. Such models would offer a more proactive defense mechanism, reducing the reliance on prior knowledge and enabling real-time adaptation to new attack strategies.

In summary, while the body of research on attack detection in SDN has yielded important results, there remain several critical areas in need of further exploration. Addressing the challenges of feature selection, broadening the scope beyond DDoS attacks, incorporating multi-flow analysis, and building models capable of detecting unseen attacks are essential steps toward advancing security measures in SDN environments.

## 2.2 Background Algorithm of BERT-base-uncased

The BERT-base-uncased model represents a significant advancement in natural language processing (NLP) by offering a robust framework for understanding textual data through its bidirectional transformer architecture (Geetha and Renuka, 2021). This model captures dependencies and semantic relationships in both directions within a sequence, enabling deep contextual learning (Devlin et al. 2018). In the context of attack detection in Software-Defined Networking (SDN) environments, BERT-base-uncased provides a unique ability to model network traffic as text, making it well-suited for analyzing patterns and identifying a wide range of attacks.

One of the key strengths of BERT-base-uncased is that it is pre-trained on vast amounts of textual data, including the entirety of the English Wikipedia and the BookCorpus dataset. This pre-training enables the model to learn rich representations of language and context, allowing it to transfer this knowledge effectively to other tasks, such as attack detection in SDN. By being exposed to diverse data during pre-training, BERT develops a deep understanding of patterns, which makes it especially effective when applied to new domains like network security, where it can generalize well across different types of network traffic.

At its core, BERT-base-uncased utilizes a multi-layer transformer architecture with 12 transformer encoder layers. Each layer incorporates two essential components: multi-head self-attention and feedforward neural networks. This architecture allows BERT to capture complex relationships between tokens. Bidirectional processing is crucial in comprehending not only the direct sequence of traffic flows but also latent dependencies that may signal attacks. For example, in SDN environments, the model's ability to focus on the relationships between network flows is a key advantage in identifying malicious behavior.

The process begins with tokenization, where input text (or traffic data, in the case of SDN) is divided into subword units using WordPiece tokenization. Each token is transformed into a 768-dimensional embedding that captures both semantic and syntactic information. These embeddings are crucial for effectively representing features of network traffic in a way BERT can understand. Segment embeddings are added to distinguish different sequences within the data, while positional embeddings ensure that BERT recognizes the order of tokens, which is important for understanding traffic flows over time. The transformation of a token into its corresponding embedding vector is represented as the formula (1).

$$E_t = \text{Embed}(t) \qquad (1)$$

Once tokenized, the data passes through transformer layers, where the multi-head self-attention mechanism plays a pivotal role. This mechanism allows each token to consider all other tokens in the sequence, capturing intricate relationships and dependencies between them. For traffic data in SDN, where the relationships between different flows may reveal patterns indicative of attacks, this mechanism is crucial. The self-attention mechanism can be mathematically described as the formula (2).

$$\text{Attention}(Q, K, V) = \text{softmax}\left(\frac{QK^T}{\sqrt{d_k}}\right)V \qquad (2)$$

Following the self-attention layer, the data is processed by a feedforward neural network, which applies non-linear transformations to the extracted features, further refining the understanding of the input data. This is essential for deriving high-level patterns in network traffic that might signal anomalous behavior or potential attacks. The feedforward neural network layer can be described by the following formula (3).

$$\text{FFN}(x) = \text{ReLU}(xW_1 + b_1)W_2 + b_2 \qquad (3)$$

The pre-training phase of BERT involves two key objectives: Masked Language Modeling (MLM) and Next Sentence Prediction (NSP). In MLM, random tokens in the input sequence are masked, and the model is trained to predict these tokens using the surrounding context. This allows BERT to learn the contextual relationships within the data, which is critical for generalizing across various network traffic patterns and detecting attacks in SDN. The objective function for MLM is expressed as formula (4).

$$\mathcal{L}_{\text{MLM}} = \sum_{t=1}^{T} \log P(t|\text{context}) \qquad (4)$$

Next Sentence Prediction (NSP) helps BERT understand relationships between sequences by predicting whether two given sentences are adjacent. In an SDN context, this capability enhances BERT's ability to model continuous sequences of traffic, helping detect patterns that deviate from the norm and signal potential security threats. The objective function for NSP is represented as formula (5).

$$\mathcal{L}_{\text{NSP}} = -\frac{1}{N}\sum_{i=1}^{N} [y_i \log P(y_i) + (1 - y_i)\log(1 - P(y_i))] \qquad (5)$$

Once pre-training is completed, BERT-base-uncased is fine-tuned for specific tasks, such as detecting a variety of attacks in SDN. Fine-tuning allows the model to adapt to the specific characteristics of network traffic, enhancing its ability to generalize to both known and previously unseen attacks. This adaptability is especially important in SDN environments, where attack patterns can evolve and vary significantly. By training on network traffic data, BERT is able to learn subtle patterns that help identify malicious activity, even when attackers employ new tactics or strategies.

In summary, the BERT-base-uncased architecture, with its bidirectional transformers, multi-head self-attention, and pre-training on vast data, offers a powerful framework for detecting attacks in SDN environments. Its capacity to capture complex relationships within data makes it well-equipped to handle the dynamic nature of network traffic and to identify both known and emerging security threats.

## 3. Proposed Methodology for Attack Detection in SDN Environments

With the rise in sophisticated cyber-attacks targeting Software-Defined Networking (SDN), there is a critical need for advanced detection methods that can efficiently identify and mitigate these threats. In response, this study presents a novel approach that leverages Natural Language Processing (NLP) techniques and the power of the BERT model to detect attacks in SDN environments. Our methodology consists of four key components as shown in Figure 2:

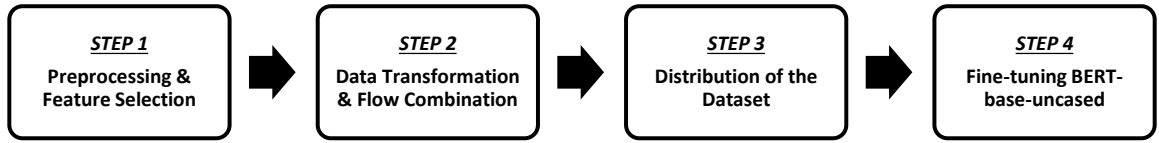

*Figure 2:Methodology Overview*

### 3.1 Data Processing and Feature Selection

Effective data processing and feature selection are essential for developing high-performance machine learning models, particularly large language models (LLMs), as they directly impact model accuracy, efficiency, and generalization. In this study, we performed a two-phase process: data preprocessing followed by feature selection, both of which were crucial for optimizing our proposed attack detection framework.

The first step in preparing the dataset for training involved rigorous preprocessing to ensure that the data was clean and suitable for the model. The InSDN dataset originally contained socket-related features such as Flow ID, Source IP, Source Port, Destination IP, Destination Port, Protocol, and Timestamp. While these features capture important network-specific details, they were removed manually because their values can vary greatly across different networks, leading to overfitting. Removing these features ensured that the model would not rely on network-specific attributes that hinder generalization. After this preprocessing step, the dataset consisted of 76 features plus the label, amounting to a total of 343,889 records. This refined dataset focused on features that are stable across various network environments, capturing essential aspects of network traffic behavior without introducing variability that could degrade model performance.

Once the dataset was preprocessed, we applied feature selection (Algorithm 1) to optimize model performance and reduce computational complexity. Using the Random Forest classifier, a reliable ensemble method that ranks features based on their contribution to decision-making, we identified the top 10 most influential features from the dataset. Random Forest evaluates feature importance by measuring how much each feature reduces uncertainty or impurity across multiple decision trees, providing a robust measure of feature relevance. The selected features, Flow Duration, Flow Pkts/s, Flow IAT Mean, Flow IAT Max, Bwd IAT Tot, Bwd IAT Mean, Bwd Header Len, Bwd Pkts/s, Pkt Len Max, Init Bwd Win Byts, were the most critical in distinguishing between different attack patterns (Table 2). This strategic selection of key features minimized noise and improved the model's focus on the most informative data, enhancing both accuracy and efficiency. By reducing the dataset to these essential features, we not only accelerated the training process but also strengthened the model's ability to generalize across different network environments, enabling it to detect both known and novel attacks with greater precision. This targeted approach was instrumental in creating a robust model that can efficiently handle large-scale datasets while maintaining adaptability to evolving cyber threats.

```
Input: A preprocessed dataset in tabular format
Output: A subset of the dataset containing only the top 10 important features
  1. Start
  2. Load dataset into a DataFrame
  3. Separate dataset into features (X) and target (y)
  4. Initialize Random Forest Classifier with fixed random state
  5. Train classifier with X and y
  6. Extract feature importances from the classifier
  7. Create DataFrame with 'Feature' and 'Importance' columns
  8. Sort DataFrame by 'Importance' in descending order
  9. Select top 10 features based on importance scores
  10. Filter dataset to include top 10 features and target variable
  11. Export filtered dataset to a new CSV file
  12. End
```
*Algorithm 1: Feature Selection Using Random Forest Classifier*

*Table 2: The best 10 Top selected features*

| # | Features | Explanation | Importance Value |
|---|---|---|---|
| 1 | Flow Duration | The total duration of the flow, from start to end, in microseconds. | 0.0456 |
| 2 | Flow Pkts/s | Number of packets transmitted per second in the flow. | 0.0375 |
| 3 | Flow IAT Mean | Mean time between packets in a flow, indicating average packet inter-arrival time. | 0.0398 |
| 4 | Flow IAT Max | Maximum time between two consecutive packets in the flow. | 0.0370 |
| 5 | Bwd IAT Tot | Total time between all packets in the backward direction during the flow. | 0.0309 |
| 6 | Bwd IAT Mean | Mean time between packets in the backward direction. | 0.0345 |
| 7 | Bwd Header Len | The total length of the backward packet headers in the flow. | 0.1018 |
| 8 | Bwd Pkts/s | Rate of packets being sent in the backward direction per second. | 0.0550 |
| 9 | Pkt Len Max | Maximum packet length observed during the flow. | 0.0354 |
| 10 | Init Bwd Win Byts | Initial backward window size in bytes during the connection. | 0.1052 |

## 3.2 Transformed Dataset to NLP and Flow Combination

In this study, we leveraged the capabilities of Natural Language Processing (NLP) and Large Language Models (LLMs) to address the challenge of attack detection in SDN environments. One of the novel approaches in SDN security we employed was the transformation of the dataset's selected features into a format suitable for NLP processing (Algorithm 2). Specifically, we converted the network traffic flow data into natural language sentences as shown in (Table 3). This transformation enabled us to utilize advanced language models like BERT, which are designed to process textual data, to interpret and detect attack patterns embedded within network traffic. The key advantage of this transformation lies in its ability to exploit BERT's proficiency in capturing contextual relationships, allowing the model to better identify complex attack patterns that might be overlooked by conventional ML and DL detection methods. By framing network traffic as structured sentences, we enriched the data representation, which enhanced the model's capacity to distinguish between benign traffic and malicious activities, thus contributing to more accurate and effective detection of diverse attack types.

| | |
|---|---|
| Input: A dataset containing selected features and labels | |
| Output: A dataset containing Sentence and labels | |
| 1. Start | |
| 2. Load the dataset into a DataFrame | |
| 3. Clean column names by stripping any leading or trailing spaces | |
| 4. Define a function to format and combine feature names and values into a single text string | |
| 5. Apply the function to each row to create a 'Sentence' column | |
| 6. Extract the 'text' and 'label' columns into a new DataFrame | |
| 7. Save the new DataFrame to a CSV file | |
| 8. End | |

*Algorithm 2: Transform Dataset for Natural Language Processing*

*Table 3: A sample of Flow Data into Natural Language Processing Sentences*

| Sentence | label |
|---|---|
| Flow Duration=1605449, Flow Pkts/s=159.4569494, Flow IAT Mean=6295.878431, Flow IAT Max=859760, Bwd IAT Tot=1603130, Bwd IAT Mean=10831.95946, Bwd Header Len=3004, Bwd Pkts/s=92.8089276, Pkt Len Max=27300, Init Bwd Win Byts=64240 | 0 |
| Flow Duration=63199059, Flow Pkts/s=0.158230204, Flow IAT Mean=7022117.667, Flow IAT Max=63200000, Bwd IAT Tot=63200000, Bwd IAT Mean=10500000, Bwd Header Len=232, Bwd Pkts/s=0.110761143, Pkt Len Max=30, Init Bwd Win Byts=63 | 1 |

After merely transforming individual network flows into sentences, we also introduced a novel flow combination method to strengthen the model's decision-making process (Algorithm 3). Instead of making the attack decision for each flow in isolation, which may limit the model's contextual understanding, we combined every four consecutive flows into a single sentence as shown in (Table 4) to make a robust attack decision. This approach provided the model with a broader view of network traffic patterns, facilitating a more comprehensive analysis of evolving threats. The label of each combined sentence was determined by the label of the fourth flow in the sequence, ensuring that the model learned from the most recent behavior within the aggregated flows. By analyzing multiple flows at once, the model was able to capture both immediate and latent indicators of attacks, which are often spread out over several flows. This combination of flows allowed for deeper insight into traffic behavior, enhancing the model's ability to detect sophisticated, multi-stage attacks that often evade detection when viewed in isolation.

The flow combination strategy also contributed to improved accuracy and a reduction in false positives by offering a more holistic perspective on network behavior. In conventional single-flow analysis, the model's decisions are based on a limited snapshot of network activity, which can lead to misclassification or overlook gradual attack patterns. By incorporating information from multiple flows, our approach provided the model with richer context, making it more adept at identifying subtle anomalies and distinguishing between normal traffic fluctuations and genuine threats. This contextual awareness is particularly beneficial for detecting complex attacks, such as advanced persistent threats (APTs) or slow-moving DDoS attacks, where malicious activities unfold incrementally across multiple flows.

Furthermore, the combination of flows enhanced the model's robustness in environments with hardware limitations. Processing each flow individually can introduce a significant computational burden, especially in resource-constrained settings. By combining flows, we reduced the overall volume of data the model had to process without sacrificing the quality of information. This reduction in computational complexity allowed the model to make more efficient predictions, even in systems with limited processing power. At the same time, the flow combination method filtered out unnecessary noise and anomalies that may arise from analyzing isolated traffic flows, ensuring that the model remained focused on the most critical aspects of the network's behavior. This balance between computational efficiency and decision-making accuracy makes our approach particularly suited for real-world SDN applications, where rapid and reliable attack detection is essential despite hardware constraints.

---

Input: A dataset containing sentences and label
Output: A dataset where every 4 sentences are combined into one sentence
1. Start
2. Load the dataset into a DataFrame
3. Initialize an empty list to store the combined sentences
4. For every set of 4 consecutive sentences in the DataFrame:
   a. Combine the 4 consecutive sentences into a single new sentence
   b. Take the label from the last sentence in the set
   c. Append the combined new sentence and label to the list
5. Convert the list of combined sentences and labels back into a DataFrame
6. Save the new DataFrame to a CSV file
7. End

*Algorithm 3: Flows Combination*

*Table 4: A sample of Combined flows into a Single Sentence*

| Sentence | label |
|---|---|
| Flow Duration=4186, Flow Pkts/s=955.566173, Flow IAT Mean=1395.333333, Flow IAT Max=3094, Bwd IAT Tot=1027, Bwd IAT Mean=1027, Bwd Header Len=40, Bwd Pkts/s=477.7830865, Pkt Len Max=46, Init Bwd Win Byts=64240 Flow Duration=3849, Flow Pkts/s=779.4232268, Flow IAT Mean=1924.5, Flow IAT Max=3690, Bwd IAT Tot=0, Bwd IAT Mean=0, Bwd Header Len=20, Bwd Pkts/s=259.8077423, Pkt Len Max=0, Init Bwd Win Byts=64240 Flow Duration=3082, Flow Pkts/s=973.3939001, Flow IAT Mean=1541, Flow IAT Max=3033, Bwd IAT Tot=0, Bwd IAT Mean=0, Bwd Header Len=20, Bwd Pkts/s=324.4646334, Pkt Len Max=31, Init Bwd Win Byts=64240 Flow Duration=4708, Flow Pkts/s=637.213254, Flow IAT Mean=2354, Flow IAT Max=4527, Bwd IAT Tot=0, Bwd IAT Mean=0, Bwd Header Len=20, Bwd Pkts/s=212.404418, Pkt Len Max=0, Init Bwd Win Byts=64240 | 0 |
| Flow Duration=30, Flow Pkts/s=66666.66667, Flow IAT Mean=30, Flow IAT Max=30, Bwd IAT Tot=30, Bwd IAT Mean=30, Bwd Header Len=0, Bwd Pkts/s=66666.66667, Pkt Len Max=0, Init Bwd Win Byts=-1 Flow Duration=16, Flow Pkts/s=125000, Flow IAT Mean=16, Flow IAT Max=16, Bwd IAT Tot=16, Bwd IAT Mean=16, Bwd Header Len=0, Bwd Pkts/s=125000, Pkt Len Max=0, Init Bwd Win Byts=-1 Flow Duration=16, Flow Pkts/s=125000, Flow IAT Mean=16, Flow IAT Max=16, Bwd IAT Tot=16, Bwd IAT Mean=16, Bwd Header Len=0, Bwd Pkts/s=125000, Pkt Len Max=0, Init Bwd Win Byts=-1 Flow Duration=18, Flow Pkts/s=111111.1111, Flow IAT Mean=18, Flow IAT Max=18, Bwd IAT Tot=18, Bwd IAT Mean=18, Bwd Header Len=0, Bwd Pkts/s=111111.1111, Pkt Len Max=0, Init Bwd Win Byts=-1 | 1 |

### 3.3 Distribution of the Dataset

The dataset used in this study consists of two main classes: normal traffic, labeled as 0, and various types of attack traffic, labeled as 1. After combining network flows, the resulting dataset contains 17,106 sentences representing normal traffic and 68,867 sentences representing attack traffic. The attack traffic spans categories such as DDoS, Probe, DoS, Brute Force Attack (BFA), Web-based attacks, Botnet activities, and User to Root (U2R) attacks. This clear distinction between normal and attack traffic is crucial for training the BERT model to accurately classify and distinguish between benign and malicious network behaviors.

To comprehensively evaluate the performance of our proposed BERT model in detecting SDN network threats, we design two experimental scenarios. In the first scenario, the model is trained and tested on datasets that include normal traffic as well as all types of attack traffic. This scenario allows us to assess the model's ability to generalize across different attack types when the training data includes examples from each class. Specifically, the training set consists of 11,632 sentences of normal traffic and 46,829 sentences of attack traffic including DDoS, Probe, DoS, BFA, Web, BOTNET, and U2R. The validation set includes 2,053 sentences of normal traffic and 8,264 attack sentences, while the test set contains 3,421 normal sentences and 13,774 attack sentences. This balanced distribution across the training, validation, and testing phases ensures a comprehensive evaluation of the model's performance on both normal and malicious traffic. The details of the First scenario distribution are summarized in Table 5.

In the second scenario, we investigate the model's ability to detect unseen attacks. To simulate this, the model is trained on a dataset that includes only normal traffic and two attack types—DDoS and DoS. The training set for this scenario consists of 14,626 normal samples (labeled as 0) and 37,525 attack samples (labeled as 1) from DDoS and DoS. The validation set contains 1,625 normal samples and 4,170 attack samples from the same two categories. The test set is designed to challenge the model by including not only normal traffic but also a variety of attack types that were not seen during the training process. This test set contains 855 normal samples and 27,172 attack samples, covering seven attack categories: DDoS, Probe, DoS, BFA, Web, BOTNET, and U2R. By evaluating the model in this second scenario, we aim to measure its resilience and adaptability in handling previously unseen types of attacks, which is crucial for real-world applications where new, evolving attack patterns are constantly emerging. The details of the Second scenario distribution are summarized in Table 5.

*Table 5: Distribution of Dataset*

| Scenario | Train Dataset | | Validation Dataset | | Test Dataset | |
|---|---|---|---|---|---|---|
| | Sentences | Type of Traffic | Sentences | Type of Traffic | Sentences | Type of Traffic |
| First scenario | 11,632 | Normal | 2,053 | Normal | 3,421 | Normal |
| | 46,829 | DDoS, Probe, DoS, BFA, Web, BOTNET, U2R | 8,264 | DDoS, Probe, DoS, BFA, Web, BOTNET, U2R | 13,774 | DDoS, Probe, DoS, BFA, Web, BOTNET, U2R |
| | 58,461 | | 10,317 | | 17,195 | |
| Second scenario | 14,626 | Normal | 1,625 | Normal | 855 | Normal |
| | 37,525 | DDoS, DoS | 4,170 | DDoS, DoS | 27,172 | DDoS, Probe, DoS, BFA, Web, BOTNET, U2R |
| | 52,151 | | 5,795 | | 28,027 | |

### 3.4 Fine-tuning Process of the BERT-base-uncased Model

The decision to use the pre-trained BERT-base-uncased model was a deliberate and strategic choice, grounded in its exceptional track record in sequence classification tasks. BERT (Bidirectional Encoder Representations from Transformers) has revolutionized natural language processing (NLP) by introducing a bidirectional framework that enables the model to understand the context of words by looking at them in both forward and backward directions simultaneously. This capability is useful for detecting attacks in SDN environments, where the sequence of network packet flows often harbors subtle patterns indicative of malicious activity. Leveraging the pre-trained BERT-base-uncased model, which has been fine-tuned on an extensive corpus of text data, allowed us to transfer its robust language comprehension into the domain of network traffic analysis. By capitalizing on BERT's ability to discern nuanced patterns and relationships in sequential data, we positioned the model to excel in identifying both known and emerging threats within SDN, significantly enhancing attack detection.

Fine-tuning BERT for SDN attack detection involved a systematic process (figure 3) of adapting its pre-learned language representations to the unique characteristics of our transformed network flows dataset. The representation of flows as text sequences allowed us to harness BERT's advanced NLP capabilities. This fine-tuning process refined BERT's ability to capture the subtleties of SDN traffic flows, including the relationships between benign and malicious packets, and the patterns associated with various attack types. This customization extended BERT's understanding beyond general text corpora, which typically lack the terminologies and traffic patterns specific to network security data. By tailoring the pre-trained model to the domain of network traffic, we enhanced its ability to identify sophisticated and subtle anomalies, such as low-frequency DDoS or botnet activities, that often go unnoticed by less context-aware models.

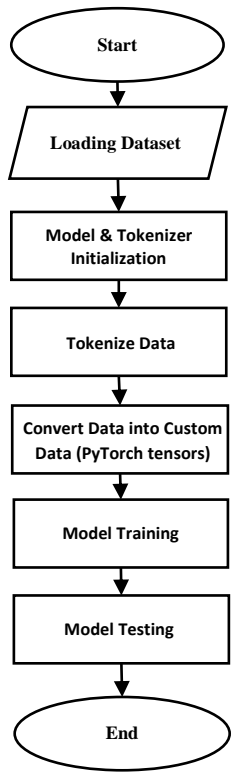

*Figure 3: Flowchart of Fine-tuning Process of the BERT-base-uncased Model*

Tokenization was a critical preparatory step in this process, ensuring that raw text data was transformed into a format BERT could process. We utilized the BERT-base-uncased tokenizer, designed to convert raw text into token IDs suitable for BERT's input layer. This tokenizer performed several key functions: it split the text into tokens, assigned each token an ID from the model's vocabulary, and handled padding and truncation to maintain uniform sequence lengths across all inputs. Our chosen maximum sequence length was 512 tokens, which is the limit for BERT, and any shorter sequences were padded with zeros to ensure batch processing efficiency. Moreover, special tokens, including the [CLS] (classification) token and [SEP] (separator) token, were added to each input sequence. The [CLS] token was essential for guiding BERT during the classification task, while the [SEP] token facilitated the distinction between multiple segments of text, if present. This comprehensive tokenization process ensured that each input was in a structured, uniform format, ready for efficient processing by BERT.

After tokenization, the next step was to structure the data into a custom dataset using PyTorch. Each dataset incorporated not only the tokenized sequences but also attention masks and corresponding labels. Attention masks played a crucial role in highlighting which tokens should be attended to by the model, improving its ability to focus on the most relevant portions of the input data. This careful structuring ensured consistency across all inputs, enabling the model to learn effectively from both normal and anomalous network traffic patterns. By leveraging PyTorch, we ensured efficient dataset handling, facilitating seamless integration with the BERT model for both training and evaluation phases.

The fine-tuning process was further optimized by configuring the TrainingArguments. These arguments were meticulously selected to strike a balance between training efficiency and model performance. A learning rate of 1e-5 was chosen to ensure stable weight updates, while a weight decay of 0.01 was implemented to prevent overfitting. The batch size was set to 128 per device, allowing for the processing of sufficient data per step without exhausting GPU memory. Gradient accumulation steps were set to 1 for quicker convergence, and training epochs were limited to 1 to prioritize frequent evaluations over prolonged training. The evaluation and save strategies were configured to trigger every 100 steps, ensuring that the model's performance was regularly assessed and that the best-performing models were saved for later use. Additionally, early stopping was implemented to halt training once no improvement was observed, thus reducing unnecessary computational expenditure. The AdamW optimizer and cross-entropy loss function were used by default, providing efficient weight updates and a suitable loss metric for binary classification. The inclusion of mixed precision training (fp16=True) further accelerated training by leveraging GPU capabilities without sacrificing model accuracy. Finally, warmup steps of 500 were incorporated to prevent drastic updates during the initial training phase, ensuring a more stable and gradual learning curve. All the key parameters used in the fine-tuning process are summarized in Table 6.

*Table 6: The Parameters of Training Arguments*

| Parameter | Value | Explanation |
|---|---|---|
| Learning Rate | 1e-5 | Ensures stable and gradual weight updates during training. |
| Weight Decay | 0.01 | Prevents overfitting by regularizing model weights. |
| Batch Size | 128 | Processes a sufficient amount of data per step while avoiding GPU memory exhaustion. |
| Gradient Accumulation | 1 | Accelerates convergence by updating gradients after each batch. |
| Epochs | 1 | Limits training to prioritize frequent evaluation rather than prolonged training. |
| Evaluation & Save Steps | Every 100 steps | Ensures regular model performance assessment and saves the best models during training. |
| Early Stopping | Enabled | Halts training when no improvement is detected, reducing unnecessary computational resources. |
| Optimizer | AdamW | Provides efficient weight updates tailored for modern deep learning. |
| Loss Function | Cross-Entropy | Suitable for binary classification, and measuring prediction error. |
| Mixed Precision (fp16) | TRUE | Leverages GPU capabilities to speed up training without sacrificing accuracy. |
| Warmup Steps | 500 | Gradually increase the learning rate during the early training phase to prevent drastic updates. |

The entire training process was streamlined using the Hugging Face Trainer class, which automated key aspects of training, such as model optimization, learning rate scheduling, and evaluation. This framework was crucial for maintaining an efficient and organized training pipeline. One of the key advantages of using the Trainer was its ability to handle logging, metric computation, and model checkpointing automatically. Learning rate scheduling was another integral feature, as it dynamically adjusted the learning rate throughout training, facilitating smoother convergence and preventing overshooting. The Trainer's efficient handling of model training, combined with our carefully selected parameters, ensured that the fine-tuning process was both robust and reliable, resulting in a high-performing model capable of accurately detecting SDN attacks.

## 4. Experimental Performance analysis and results

In this section, we provide a comprehensive analysis of the experimental evaluation conducted to assess the effectiveness of the proposed approach. We detail the experimental setup and environment, describe the selected dataset, and discuss the results of feature importance. Additionally, we outline the performance metrics employed to evaluate the model's capabilities. Furthermore, we present an in-depth analysis of the model's performance, highlighting its ability to detect attacks in SDN environments. This analysis includes a comparison of our model with traditional DNN and CNN architectures, illustrating the advantages of our approach in terms of accuracy, precision, and overall effectiveness in attack detection.

### 4.1 Experimental Setup Environment

To ensure optimal performance and reliability for our research, we conducted experiments on a high-performance workstation designed to handle the computational demands of advanced machine learning tasks. Table 7 summarizes the hardware and software configurations employed throughout the study. The workstation operates on Ubuntu 22.04 with Linux Kernel version 6.5.0-44-generic, featuring an AMD EPYC 9654 processor with 96 cores and 192 threads, running at clock speeds between 1.50 and 3.70 GHz. This powerful processor is paired with 503 GiB of RAM to facilitate smooth data processing and large-scale simulations. For GPU acceleration, we utilized an NVIDIA A100 80GB PCIe GPU, which supports CUDA version 12.4 and is powered by the NVIDIA driver version 550.90.07, providing exceptional parallel processing capabilities for training large language models. Python 3.10.12 was selected as the programming language, while Jupyter Notebook served as the primary development environment, offering an intuitive platform for iterative experimentation. This robust configuration enabled the efficient execution of complex algorithms and ensured that our research was supported by the highest levels of computational performance.

*Table 7: Experimental Setup Environment*

| Component | Details |
|---|---|
| Workstation | Ubuntu 22.04, Linux Kernel 6.5.0-44-generic |
| CPU | AMD EPYC 9654, 96 cores (192 threads), 1.50 - 3.70 GHz |
| RAM | 503 GiB |
| GPU | NVIDIA A100 80GB PCIe, Driver 550.90.07, CUDA 12.4 |
| Python version | 3.10.12 |
| IDE Platform | Jupyter Notebook |

### 4.2 Dataset Selection and Results of Features Importance

The selection of a robust and comprehensive dataset is paramount for assessing the effectiveness of large language models, particularly in the context of SDN security. For this study, we employed the InSDN (Intrusion Detection in Software-Defined Networking) dataset (Elsayed et al., 2020), a benchmark specifically designed to capture a wide range of traffic patterns and attack scenarios within SDN environments (available at InSDN). The InSDN dataset stands out due to its extensive coverage of both normal and malicious traffic, encompassing various attack types, including Distributed Denial of Service

(DDoS), Denial of Service (DoS), Probe, Brute Force Attacks (BFA), Web Attacks, BOTNET activities, and User-to-Root (U2R) attacks. This broad spectrum of attack scenarios ensures a thorough evaluation of the model's capability to generalize across different threats. With 68,424 flows of normal traffic and 275,465 flows of attack traffic, the InSDN dataset provides a robust foundation for developing and testing advanced SDN security solutions. Furthermore, its comprehensive feature set of 84 distinct attributes offers detailed insights into network traffic behaviors and patterns. Figure 4 illustrates the distribution of these traffic types within the dataset, reinforcing the InSDN dataset's suitability for developing large language models that are both effective and adaptable in the dynamic landscape of SDN security.

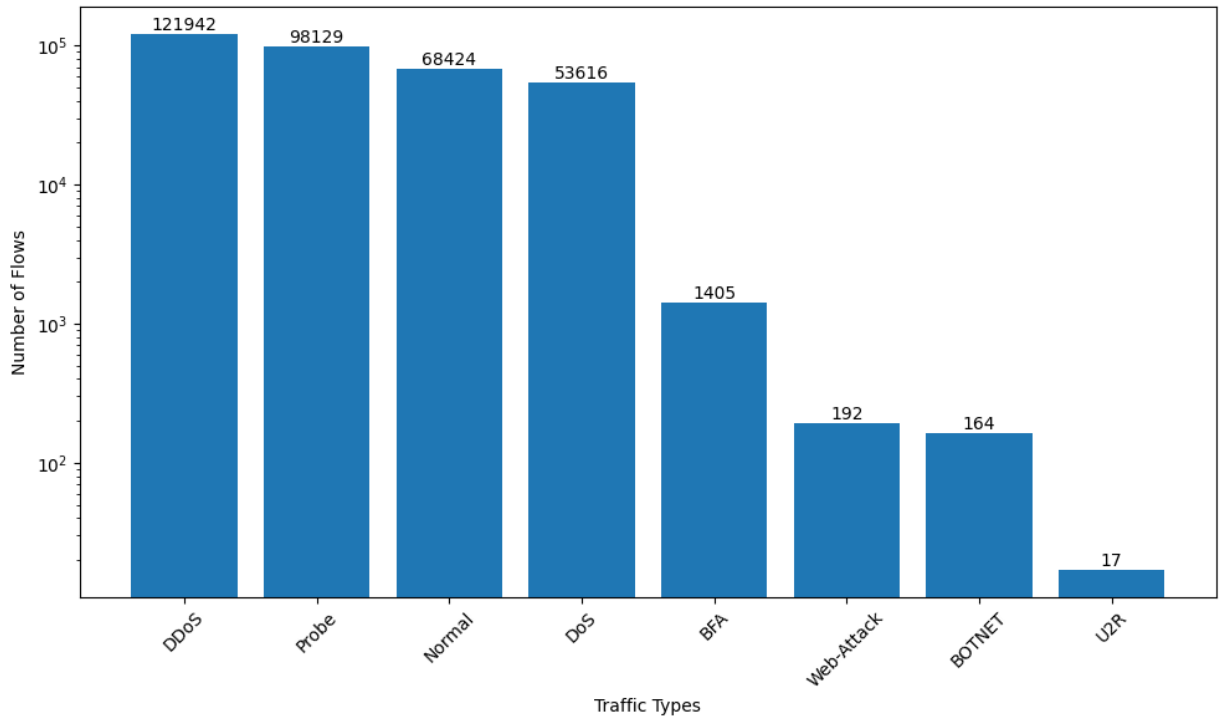

*Figure 4: Distribution of Traffic Types in InSDN Dataset*

As a result of our Algorithm 1, which applies a feature selection method to identify the top 10 most impactful features, we present the feature importance analysis generated by the Random Forest classifier (Figure 5)." This figure highlights the relative importance of each feature based on its contribution to classification performance. The features at the top of the chart, such as Init Bwd Win Byts, Bwd Header Len, Bwd Pkts/s, Flow Duration, Flow IAT Mean, Flow Pkts/s, Flow IAT Max, Pkt Len Max, Bwd IAT Mean, and Bwd IAT Tot demonstrate significantly higher importance values, underscoring their critical role in accurately differentiating normal traffic from various attack types. The importance scores in this figure emphasize how critical it is to focus on a select group of features that effectively capture distinctive characteristics of network traffic in SDN environments. By focusing on these high-importance features, our model achieves both enhanced performance and reduced computational complexity, which is essential for efficient, scalable deployment in real-world networks.

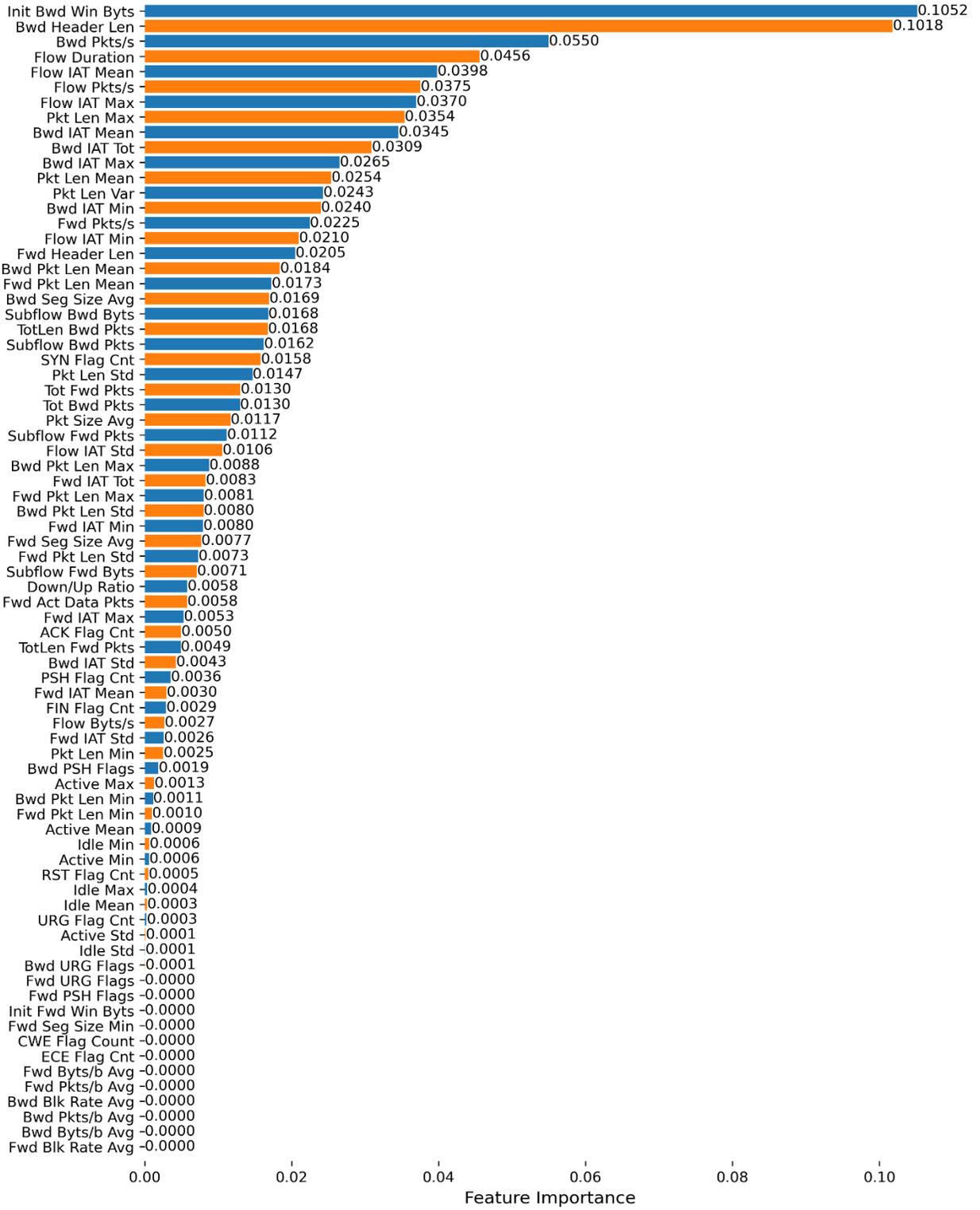

Figure 5: Results of Features Importance

### 4.3 Performance Metrics

To thoroughly assess the performance of our proposed model, we employed a set of widely accepted evaluation metrics: Accuracy, Precision, Recall, and F1-score, along with the Receiver Operating Characteristic (ROC) curve and the Area Under the Curve (AUC). These metrics offer a comprehensive perspective on the model's predictive capabilities, providing deep insights into its strengths and weaknesses across different attack scenarios.

- Accuracy measures the proportion of correct predictions among all predictions, offering a broad indicator of the model's overall performance. Accuracy is calculated using the formula (6):

$$Accuracy = \frac{TP + TN}{TP + TN + FP + FN} \quad (6)$$

where TP represents true positives, TN represents true negatives, FP represents false positives, and FN represents false negatives.

- Precision focuses on the accuracy of positive predictions, indicating the proportion of true positive predictions among all positive predictions made by the model. It is a critical metric when the cost of false positives is high. Precision is calculated using the formula (7):

$$Precision = \frac{TP}{TP + FP} \quad (7)$$

- Recall, also known as Sensitivity or True Positive Rate, measures the model's ability to identify all relevant positive instances within the dataset. It indicates the proportion of true positive predictions among all actual positive instances. Recall is particularly important in scenarios where missing positive instances has significant consequences. The formula (8) for Recall is:

$$Recall = \frac{TP}{TP + FN} \quad (8)$$

- F1-score is the harmonic mean of Precision and Recall, providing a single metric that balances the trade-off between the two. It is especially useful when there is an uneven class distribution, or when both false positives and false negatives are costly. The F1-Score is computed using Equation (9):

$$F1 - score = 2 * \frac{Precision * Recall}{Precision + Recall} \quad (9)$$

- ROC Curve (Receiver Operating Characteristic) is a graphical representation that illustrates the trade-off between the True Positive Rate (Recall) and the False Positive Rate (FPR) across different threshold settings. The ROC curve provides insight into how well the model can distinguish between positive and negative classes. A curve that approaches the upper-left corner of the plot indicates better performance, as it suggests high recall with a low false positive rate.

- AUC (Area Under the ROC Curve) offers a single scalar value that quantifies the overall performance of the model based on the ROC curve. An AUC score of 0.5 indicates random guessing, while a score closer to 1.0 reflects a strong ability to discriminate between positive and negative instances. A higher AUC suggests that the model performs well across various threshold settings.

By leveraging these comprehensive evaluation metrics Accuracy, Precision, Recall, F1-Score, ROC, and AUC we ensured a holistic assessment of the model's performance. Each metric provides unique insights, allowing us to understand how well the model can distinguish between normal and malicious traffic across various attack types. This thorough evaluation enables us to fine-tune and optimize the model for real-world SDN environments, where accurate and reliable attack detection is paramount for maintaining network security.

### 4.4 Bert Model Performance Analysis

In this subsection, we provide a detailed analysis of the model's performance during the training and validation phases across our two scenarios. The first scenario involved training the model on a dataset that contained both normal and malicious traffic, representing various attack types. As shown in Table 8, the model demonstrated substantial improvement as the training progressed. At step 100, the training loss was 0.5246, which steadily decreased to 0.0063 by step 400, while the validation loss also decreased from 0.3908 to 0.0036 during the same period. The reduction in both training and validation losses signifies the model's capacity to learn from the data effectively without overfitting. The training loss remained slightly larger than the validation loss in the early stages of the training process. This discrepancy can occur because we evaluate the model every 100 steps rather than at the end of each epoch, meaning that the training loss is calculated on partially adapted model weights within an epoch. Meanwhile, the validation loss is computed over the entire validation set at consistent intervals, leading to a more stable and lower loss value in comparison. The training and validation process was completed in 238 seconds, reflecting the efficiency of the model's convergence. The performance metrics further emphasize this trend, as accuracy improved from 80.10% at step 100 to an impressive 99.95% at step 400. Precision, recall, and F1-score all followed a similar upward trajectory, reaching nearly perfect values of 0.9995 by the final step. These results underscore the model's robust ability to classify traffic accurately and consistently, as reflected by the near-zero validation loss. The accompanying graph (Figure 6) illustrates the smooth convergence of both training and validation loss curves, highlighting the stability and reliability of the model during this phase.

*Table 8: Model Performance in First Scenario*

| Step | Training Loss | Validation Loss | Accuracy | Precision | Recall | F1-score |
|------|---------------|-----------------|----------|-----------|--------|----------|
| 100  | 0.5246        | 0.3908          | 0.8010   | 0.6416    | 0.8010 | 0.7125   |
| 200  | 0.2556        | 0.0463          | 0.9978   | 0.9978    | 0.9978 | 0.9978   |
| 300  | 0.0241        | 0.0090          | 0.9986   | 0.9986    | 0.9986 | 0.9986   |
| 400  | 0.0063        | 0.0036          | 0.9995   | 0.9995    | 0.9995 | 0.9995   |

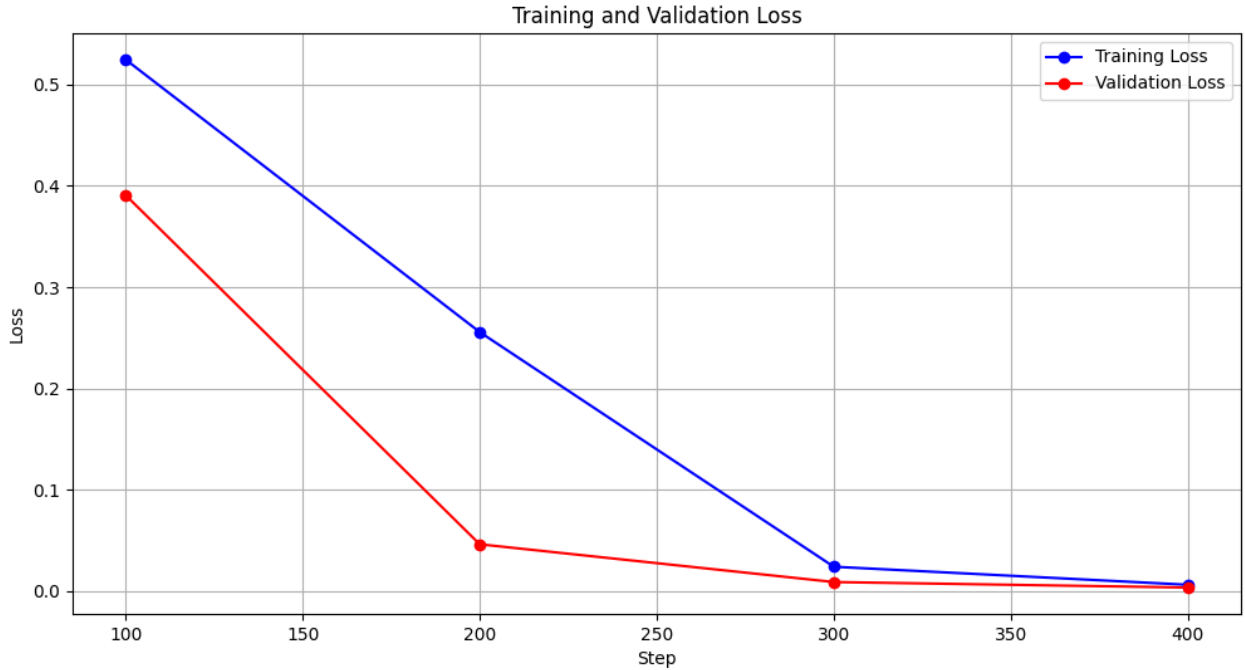

*Figure 6: Train and Validation loss in First Scenario*

In the second scenario, the model was trained under slightly different conditions, where the training data also contained a diverse mixture of normal and attack traffic. As shown in Table 9, the model's performance in this scenario followed a similar trend, with training loss decreasing from 0.5695 at step 100 to 0.0067 at step 400. Validation loss decreased from 0.4086 to 0.0044 over the same steps, further indicating that the model was generalizing well to the validation data. The training loss remained slightly larger than the validation loss in the early steps of training. This can be attributed to the fact that we evaluate the model every 100 steps, rather than at the end of each epoch, which means that the training loss is calculated on partially updated model weights within an epoch. In contrast, the validation loss is computed on the entire validation set at consistent intervals, resulting in a more stable and often lower loss compared to the training loss. The training and validation process was completed in 193 seconds, highlighting the model's efficiency in converging to optimal performance. The accuracy at the final step reached 99.89%, with precision, recall, and F1-scores all converging to approximately 0.9990. These metrics reflect the model's ability to consistently and accurately classify the validation data, which included traffic types the model had encountered during training. As seen in the graph (Figure 7), the training and validation loss curves exhibit a smooth and consistent decline, signifying the model's capacity to generalize well without signs of overfitting. This analysis demonstrates the strong performance of the model during training and validation, showcasing its ability to learn effectively from the data and maintain high accuracy across multiple steps.

*Table 9: Model Performance in Second Scenario*

| Step | Training Loss | Validation Loss | Accuracy | Precision | Recall | F1-score |
|---|---|---|---|---|---|---|
| 100 | 0.5695 | 0.4086 | 0.7976 | 0.7876 | 0.7976 | 0.7822 |
| 200 | 0.2179 | 0.0303 | 0.9979 | 0.9979 | 0.9979 | 0.9979 |
| 300 | 0.0222 | 0.0081 | 0.9990 | 0.9990 | 0.9990 | 0.9990 |
| 400 | 0.0067 | 0.0044 | 0.9990 | 0.9990 | 0.9990 | 0.9990 |

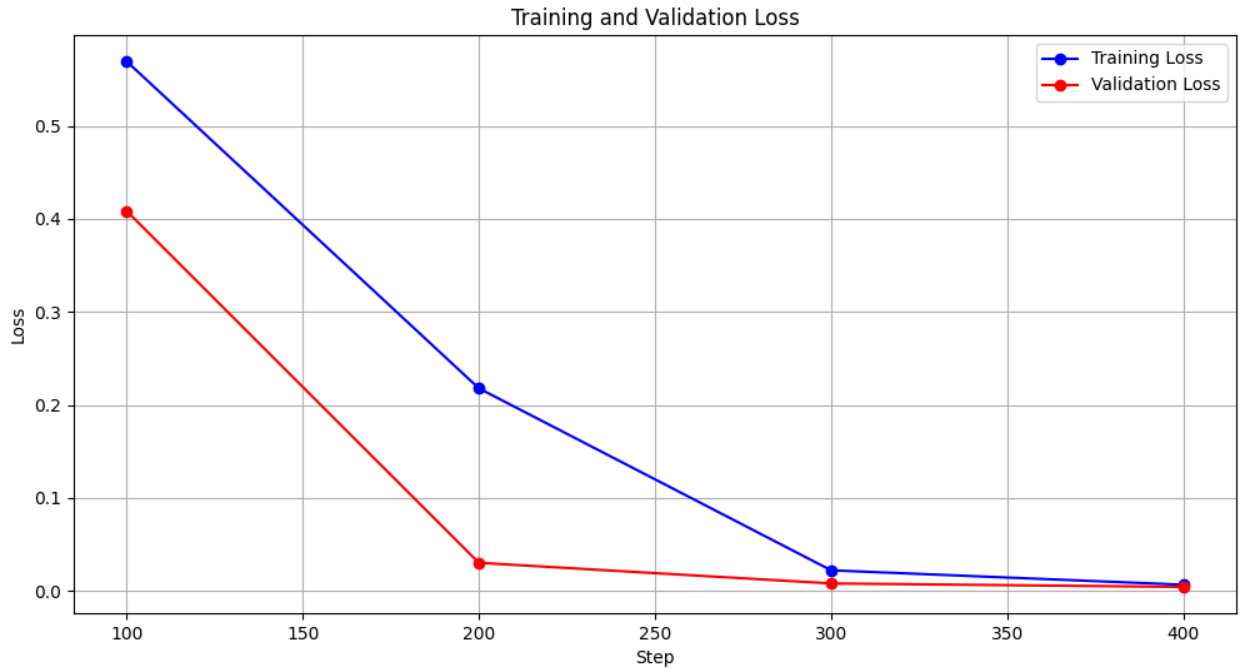

*Figure 7: Train and Validation loss in Second Scenario*

Overall, the results from both scenarios indicate that the model effectively learns from the training data, with both training and validation loss showing significant reduction and near-perfect performance metrics by the final steps. These findings reflect the robustness and efficiency of the model, as it demonstrates an excellent ability to generalize to unseen validation data without overfitting. This performance analysis is focused solely on the training and validation phases, while the final evaluation of the model on unseen test data, including the new attack types, will be presented in the subsequent results section.

### 4.5 Results of the Proposed BERT Model

To evaluate the performance and robustness of our BERT model for intrusion detection in SDN environments, we conducted experiments under two distinct scenarios. Each scenario was designed to test the model's ability to accurately detect a variety of attack types, with particular emphasis on its generalization capabilities and performance in handling unseen attacks.

In the first scenario, the model was trained and tested on datasets containing normal traffic as well as seven different types of attack traffic. The evaluation metrics reflect the model's exceptional accuracy and consistency across all performance measures. Specifically, the model achieved an accuracy of 99.96%, with precision, recall, and F1-score all at 99.96%. These metrics indicate that the model was highly effective at distinguishing between normal and malicious traffic, with minimal misclassifications. The inference process was completed in 18 seconds, further demonstrating the model's efficiency. The confusion matrix (Figure 8) for this scenario highlights the model's precision: it recorded 3,418 true negatives (correctly identified normal traffic) and 13,770 true positives (correctly identified attack traffic), with only 3 false positives and 4 false negatives. The extremely low false positive and false negative rates underscore the model's strong reliability in identifying both benign and malicious traffic. Furthermore, the ROC curve (Figure 9) for this scenario yielded an AUC of 1.00, signifying perfect separation between the two classes. This flawless performance confirms the model's ability to accurately detect a wide range of attack types.

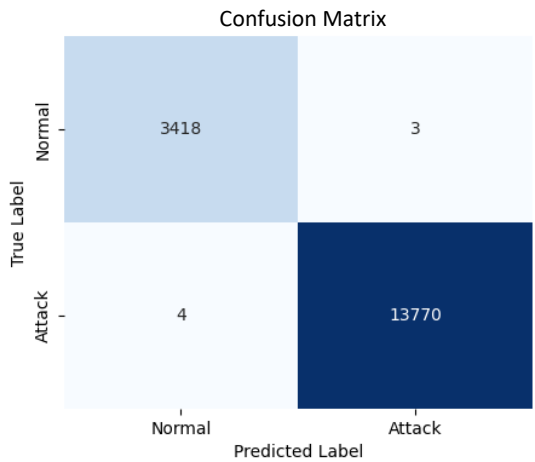 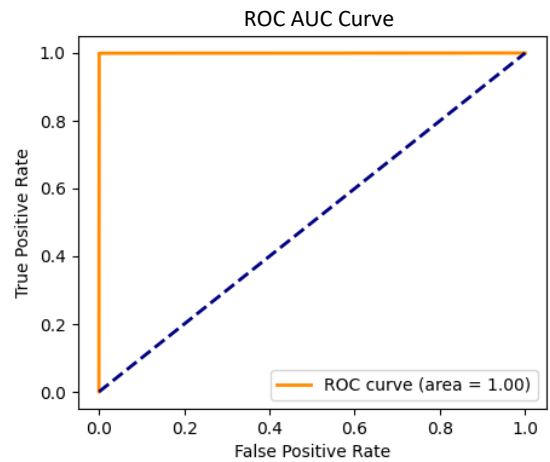

*Figure 8: Confusion Matrix Results for First Scenario*     *Figure 9: ROC AUC Curve for First Scenario*

In the second scenario, the model's ability to generalize to unseen attack types was evaluated by training it on only two types of attacks along with normal traffic and then testing it on a dataset that included seven types of attack traffic, five of which were not seen during training. This scenario was designed to simulate real-world conditions where new, previously unseen attack patterns may emerge, requiring the model to demonstrate robust adaptability. Despite this more challenging task, the model achieved an impressive accuracy of 99.96%, with precision, recall, and F1-score all at 99.96%. The inference process was completed in 29 seconds, highlighting the model's swift and effective response to previously unseen attack scenarios. The confusion matrix (Figure 10) for this scenario recorded 852 true negatives and 27,166 true positives, with only 3 false positives and 6 false negatives. The minimal number of false negatives is particularly significant in this scenario, as it demonstrates the model's capability to detect even previously unseen attack types with remarkable precision. The ROC curve (Figure 11) in this scenario also showed an AUC of 1.00, reinforcing the model's capacity to separate normal and malicious traffic even when faced with unknown threats.

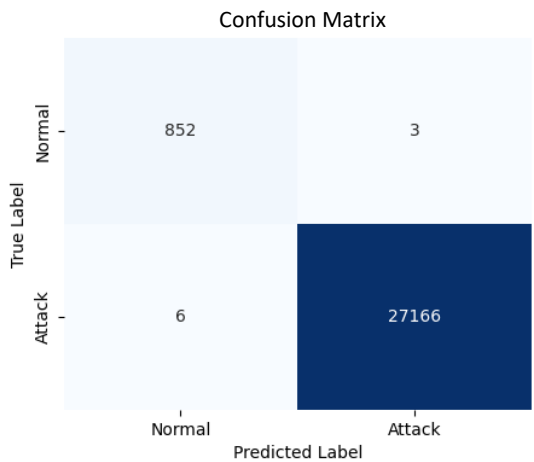 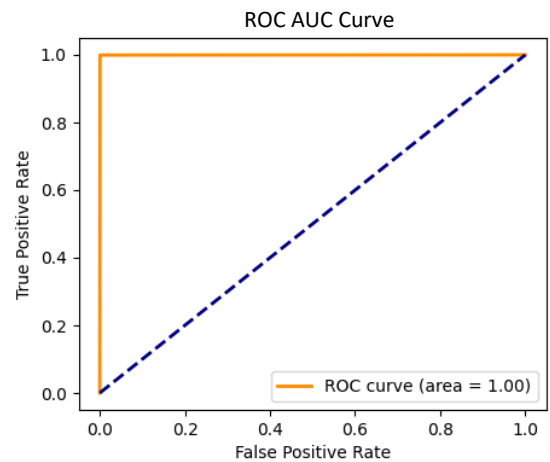

*Figure 10: Confusion Matrix Results for Second Scenario*     *Figure 11: ROC AUC Curve for Second Scenario*

The results from both scenarios demonstrate the robustness and effectiveness of the BERT-based-uncased model in detecting a wide variety of network attacks in SDN environments. The first scenario highlights the model's superior accuracy when trained and tested on all types of traffic, while the second scenario showcases its ability to generalize to unseen attacks, which is critical for real-world applications where new types of attacks frequently emerge. These findings indicate that the model not only excels in identifying known threats but is also capable of adapting to novel attack patterns, ensuring a high level of security in dynamic and evolving network environments.

### 4.6 Comparison of Our Proposed BERT Model with DNN and CNN Models

In this subsection, we compare our proposed BERT model with two commonly used neural network architectures: Deep Neural Networks (DNN) and Convolutional Neural Networks (CNN). Both models were trained and tested on the same two scenarios as our proposed BERT model, ensuring a fair and consistent evaluation of performance.

The DNN model used in this comparison consists of a simple yet effective architecture designed for binary classification. It begins with an input layer of size 512, corresponding to the number of TF-IDF features extracted from the dataset. The architecture includes two fully connected (dense) layers, each comprising 128 neurons activated by the ReLU function. To mitigate overfitting, dropout layers with a rate of 0.5 follow both dense layers. The final output layer is a single neuron with a sigmoid activation function, responsible for predicting the binary class. The architecture of the DNN model is detailed in Table 10. On the other hand, the CNN model leverages convolutional layers to capture local patterns in the data. This model comprises two 1D convolutional layers, each with 128 filters and a kernel size of 5, followed by max-pooling layers to reduce the spatial dimensions of the input. After flattening the convolutional output, a fully connected layer with 128 neurons and ReLU activation is employed, followed by a dropout layer with a rate of 0.5. The final layer, similar to the DNN model, consists of a single neuron with a sigmoid activation function. The architecture of the CNN model is also presented in Table 10. Both models DNN and CNN were compiled using the Adam optimizer and trained with the binary cross-entropy loss function. they were trained over 5 epochs, with a batch size of 128, and early stopping was implemented to prevent overfitting.

*Table 10: Architectural Details of CNN Model and CNN Model*

| Layer | DNN Architecture | CNN Architecture |
| --- | --- | --- |
| Input Layer | 512 neurons (TF-IDF feature size) | 512 input size, reshaped into (512, 1) for Conv1D |
| First Hidden Layer | 128 neurons, ReLU activation | Conv1D Layer 1: 128 filters, kernel size of 5, ReLU activation |
| Pooling Layer | N/A | Max Pooling Layer 1: Pool size of 2 |
| Second Hidden Layer | 128 neurons, ReLU activation | Conv1D Layer 2: 128 filters, kernel size of 5, ReLU activation |
| Second Pooling Layer | N/A | Max Pooling Layer 2: Pool size of 2 |
| Flatten Layer | N/A | Flatten Layer: To convert 2D output into 1D |
| Fully Connected Layer | 128 neurons, ReLU activation | 128 neurons, ReLU activation |
| Dropout Layer 1 | 0.5 dropout rate | 0.5 dropout rate |
| Dropout Layer 2 | 0.5 dropout rate | N/A |
| Output Layer | 1 neuron, Sigmoid activation for binary classification | 1 neuron, Sigmoid activation for binary classification |

**First Scenario Result: Testing on Known Attack Types**

In the first scenario, the models were trained and tested on data containing both normal traffic and seven types of attack traffic. This setup evaluates each model's effectiveness in distinguishing normal from attack traffic under conditions where all attack types are known. The results of the three models in the first scenario are summarized in Table 11.

- DNN Model: The DNN model also performed impressively, achieving 99.93% accuracy, 99.98% precision, 99.93% recall, and 99.96% F1-score, with an AUC of 1.00 (Figure 13A). It produced 2 false positives and 9 false negatives (Figure 12A). Although the DNN model's results are very close to those of the BERT model, the slightly higher number of false negatives indicates a marginally lower sensitivity to attack detection compared to BERT.

- CNN Model: The CNN model reached 99.93% accuracy, 99.97% precision, 99.94% recall, and 99.95% F1-score, with an AUC of 1.00 (Figure 13B) as well. It produced 4 false positives and 8 false negatives (Figure 12B), showing similar performance to the DNN model but with a slightly higher number of misclassifications than BERT. While the CNN model effectively distinguished normal from attack traffic, it exhibited slightly lower sensitivity in comparison to BERT.

- Proposed BERT Model: Our BERT model achieved 99.96% accuracy, precision, recall, and F1-score, with an AUC of 1.00 (Figure 13C). It generated only 3 false positives and 4 false negatives (Figure 12C), demonstrating its highly accurate and reliable performance in identifying both normal and attack traffic. This near-perfect performance highlights BERT's robustness and precision when dealing with diverse types of attack data.

While the differences between these models in the first scenario are minor, BERT's marginally better balance of false positives and false negatives, coupled with its near-perfect accuracy across all metrics, showcases its superior capacity for handling complex patterns in network traffic. Both DNN and CNN models performed extremely well, but the BERT model's self-attention mechanisms likely allowed it to capture more nuanced relationships in the dataset, resulting in fewer misclassifications.

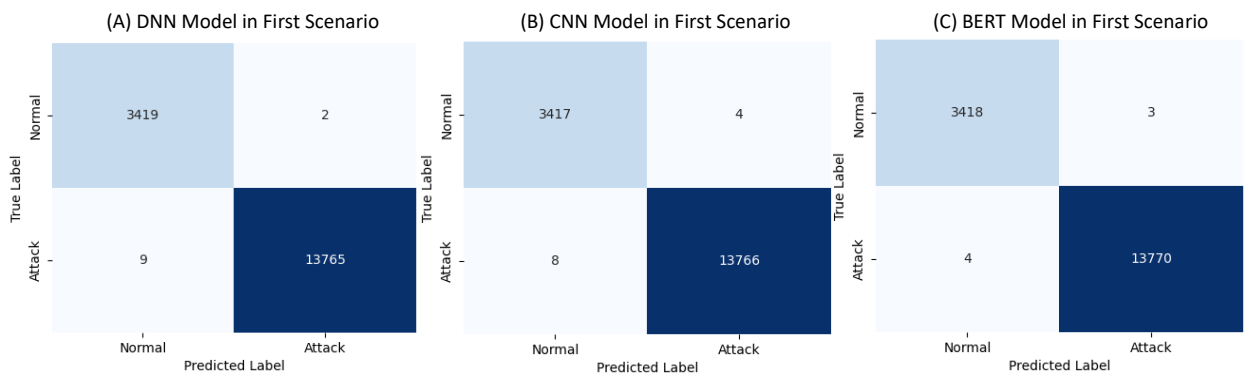

*Figure 12: Confusion Matrix Results of DNN, CNN, and BERT for First Scenario*

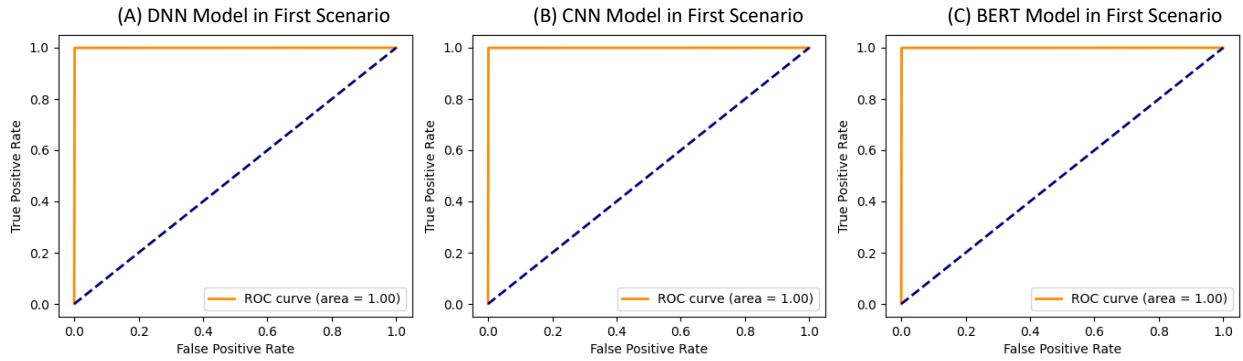

*Figure 13: ROC AUC Curve of DNN, CNN, and BERT for First Scenario*

*Table 11: Comparing Results of DNN, CNN, and BERT in the Two Scenario*

| Scenario | Model | Accuracy | Precision | Recall | F1-score |
|---|---|---|---|---|---|
| **First Scenario** | DNN | %99.93 | %99.98 | %99.93 | %99.96 |
| | CNN | %99.93 | %99.97 | %99.94 | %99.95 |
| | **Proposed BERT** | **%99.96** | **%99.96** | **%99.96** | **%99.96** |
| **Second Scenario** | DNN | %95.13 | %100 | %94.98 | %97.42 |
| | CNN | %81.66 | %100 | %81.09 | %89.55 |
| | **Proposed BERT** | **%99.96** | **%99.96** | **%99.96** | **%99.96** |

**Second Scenario Result: Testing on Unseen Attack Types**

The second scenario tests the models' ability to generalize by training them only on normal traffic and two types of attacks, then evaluating them on data that includes five additional unseen attack types. This scenario mimics real-world conditions where new types of attacks may emerge unexpectedly, making it essential to assess each model's adaptability and resilience. The results of the three models in the second scenario are summarized in Table 11.

- DNN Model: The DNN model achieved 95.13% accuracy, 94.98% recall, and 97.42% F1-score, with an AUC of 0.97 (Figure 15A). Although it maintained 100% precision with 0 false positives, it encountered a substantial number of false negatives 1,364 (Figure 14A). This suggests the DNN model has a tendency to miss certain unseen attacks, which impacts its overall reliability in adaptive detection scenarios.

- CNN Model: The CNN model showed a considerable drop in performance in this scenario, with 81.66% accuracy, 100% precision, 81.09% recall, and an F1-score of 89.55%. Its AUC dropped to 0.91 (Figure 15B), indicating a reduced ability to separate normal from attack traffic when tested on unseen attack types. The CNN model produced 5,138 false negatives (Figure 14B), which implies a high miss rate for novel attacks, despite correctly identifying normal traffic with perfect precision.

- Proposed BERT Model: The BERT model maintained excellent performance with 99.96% accuracy, precision, recall, and F1-score, and an AUC of 1.00 (Figure 15C). With only 6 false negatives and 3 false positives (Figure 14C), BERT demonstrated an outstanding ability to detect both known and unseen attacks. This result highlights BERT's ability to generalize well to novel attack patterns, making it a robust choice for adaptive network security.

This scenario underscores the BERT model's advantage in adapting to novel attacks, illustrated by its consistent performance across all metrics and its perfect AUC of 1.00. While the DNN model performed better than the CNN model in this scenario, as reflected by its AUC of 0.97, it still faced challenges with false negatives, suggesting limited generalization to unknown attacks. The CNN model's significant drop in AUC to 0.91 indicates its limited effectiveness in handling new attack types, confirming that it may not be suitable for highly dynamic security environments.

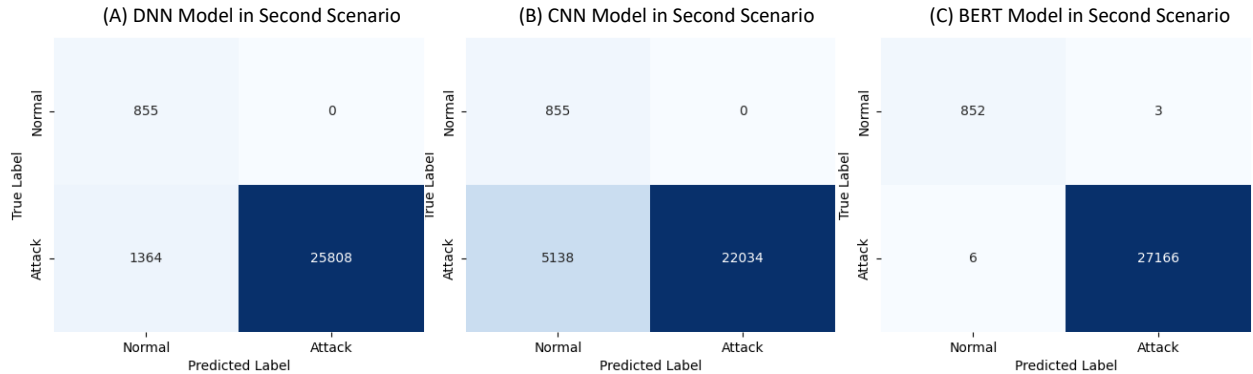

Figure 14: Confusion Matrix Results of DNN, CNN, and BERT for Second Scenario

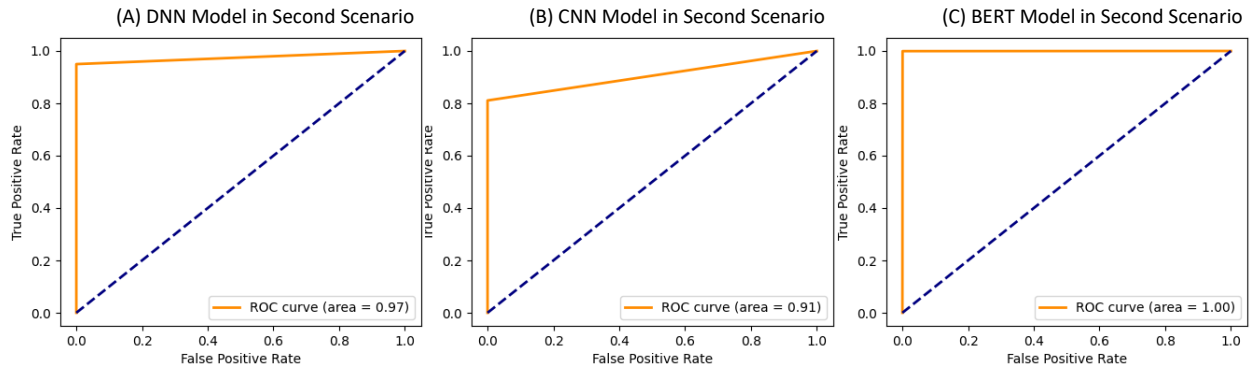

Figure 15: ROC AUC Curve of DNN, CNN, and BERT for Second Scenario

### 4.7 Comparison of Our Proposed Approach with Existing Literature

The evolving landscape of attack detection in SDN has seen significant advances through machine learning and deep learning techniques. Despite the progress, various limitations in these approaches remain unaddressed, which this research aimed to resolve. One major limitation is the over-reliance on a large number of features without the use of systematic feature selection techniques. For instance, studies like those by Najar et al. (2024), Hnamte et al. (2024), Cil et al. (2021), and Perez-Diaz et al. (2020) utilize a wide array of features that may not all contribute meaningfully to the detection task. The inclusion of redundant or non-informative features introduces the risk of overfitting, reducing the model's ability to perform effectively in real-world network environments. Our approach, in contrast, applies Random Forest-based feature selection, focusing on the most relevant and critical features. This selective process enhances the model's efficiency and significantly improves its generalizability across varied network conditions, providing a more robust solution for SDN security.

Additionally, existing literature predominantly concentrates on DDoS detection, leaving other types of cyberattacks underexplored. Najar et al. (2024), Gadallah et al. (2024), Chouhan et al. (2023), Zainudin et al. (2022), Cil et al. (2021), Perez-Diaz et al. (2020), and Elmasry et al. (2020) represent such studies, where the focus is limited to DDoS, neglecting threats such as Probe, U2R, BFA, Web Attacks, and BOTNETs. Given the growing complexity and variety of threats targeting SDN, the narrow focus on DDoS detection poses a significant gap in these approaches. Our research addresses this issue by constructing a more versatile model that can effectively detect multiple types of attacks. By expanding beyond DDoS attacks, we contribute a comprehensive SDN security solution capable of mitigating a wider array of threats.

Moreover, many of the studies reviewed, including Najar et al. (2024), Hnamte et al. (2024), Gadallah et al. (2024), Chouhan et al. (2023), Zainudin et al. (2022), Cil et al. (2021), Perez-Diaz et al. (2020), Elmasry et al. (2020) and Said Elsayed et al. (2020), rely on single flows of network traffic to make attack classification decisions. While effective to some extent, this method overlooks valuable contextual data that can be extracted from analyzing consecutive network flows. In practical, real-world settings, network interactions are complex, and the use of multiple consecutive flows for analysis would provide a richer dataset for more accurate and sophisticated attack detection. Our research incorporates multi-flow data analysis, capturing more intricate attack patterns and delivering improved detection accuracy in increasingly complex SDN environments.

Another critical shortcoming in current research is the lack of generalization to unseen attack types. None of the studies we reviewed explicitly address the challenge of detecting unknown or evolving attack patterns. Typically, models are trained and evaluated on known attack types, which limits their applicability in dynamic environments where new types of attacks can emerge. Our model differentiates itself by demonstrating the ability to handle both known and unseen attacks. By testing the model on previously unseen attack types, we have validated its robustness, highlighting its potential to offer a more adaptive and future-proof solution to SDN security challenges. The comparison of key characteristics between our approach and existing literature is summarized in Table 12.

*Table 12: Comparison of Our Proposed Approach with Existing Literature*

| Study (Year) | Feature Selection | Focus Beyond DDoS | Multi-Flow Analysis | Unseen Attack Handling |
|---|---|---|---|---|
| Najar et al. (2024) | ✘ | ✘ | ✘ | ✘ |
| Hnamte et al. (2024) | ✘ | ✓ | ✘ | ✘ |
| Gadallah et al. (2024) | ✓ | ✘ | ✘ | ✘ |
| Chouhan et al. (2023) | ✓ | ✘ | ✘ | ✘ |
| Zainudin et al. (2022) | ✓ | ✘ | ✘ | ✘ |
| Cil et al. (2021) | ✘ | ✘ | ✘ | ✘ |
| Perez-Diaz et al. (2020) | ✘ | ✘ | ✘ | ✘ |
| Elmasry et al. (2020) | ✓ | ✘ | ✘ | ✘ |
| Said Elsayed et al. (2020) | ✘ | ✓ | ✘ | ✘ |
| **Our Proposed Approach** | ✓ | ✓ | ✓ | ✓ |

## 5. Conclusion and Future Work

In this paper, we presented a comprehensive approach designed to enhance the security of SDN against persistent threats, particularly DDoS, DOS, Probe, U2R, BFA, and Web attacks. By leveraging the powerful capabilities of the pre-trained BERT-base-uncased model, our approach achieved significant improvements in attack detection performance through innovative data transformation techniques that convert network flow data into natural language, enabling BERT to effectively capture intricate patterns and relationships. Additionally, we employed robust feature selection using a Random Forest Classifier, optimizing model efficiency and performance while reducing computational overhead. The integration of previous network flow analysis further strengthened our system's ability to detect coordinated and multi-stage attacks. By considering previous flows, our framework is better equipped to identify evolving threats and respond proactively, thus enhancing the overall resilience of SDN environments.

Looking ahead, we aim to test our approach in real-world SDN environments to assess its effectiveness under practical conditions. This will involve deploying our framework across various network scenarios to evaluate its performance against actual attack vectors and diverse traffic patterns. Additionally, we plan to explore the framework's adaptability in dynamic environments, enabling it to respond effectively to changing network conditions and emerging threats. A critical area for future work is the integration of continuous learning mechanisms, allowing the model to adapt in real-time as new attack vectors are identified. This enhancement would improve the system's responsiveness and provide robust protection against evolving threats.